\documentclass[12pt]{iopart}

\usepackage{iopams}  
\usepackage{siunitx}
\usepackage{xcolor}

\usepackage{graphicx}
\graphicspath{{/Users/gcolas/Desktop/ARTICLES/YagiWg/paper-chapter-3,4}}

\begin{document}

\title{Nano antenna-assisted quantum dots emission into high-index planar waveguide}

\author{X. Yu, J.-C. Weeber, L. Markey, J. Arocas, A. Bouhelier, A. Leray, and G. {Colas des Francs}}

\address{Laboratoire Interdisciplinaire Carnot de Bourgogne (ICB), CNRS UMR 6303, Université de Bourgogne, BP 47870, 21078 Dijon, France}
\ead{gerard.colas-des-francs@u-bourgogne.fr}

\begin{abstract}
Integrated quantum photonic circuits require the efficient coupling of photon sources to photonic waveguides. Hybrid plasmonic/photonic platforms are a promising approach, taking advantage of both plasmon modal confinement for efficient coupling to a nearby emitter and photonic circuitry for optical data transfer and processing. In this work, we established directional quantum dot (QD) emission coupling to a planar  TiO$_2$ waveguide assisted by a Yagi-Uda antenna. Antenna on waveguide is first  designed by scaling radio frequency dimensions to nano-optics, taking into account the hybrid plasmonic/photonic platform. Design is then optimized by full numerical simulations. 
We fabricate the antenna  on a TiO$_2$ planar waveguide and deposit a few QDs close to the Yagi-Uda antenna. The optical characterization shows clear directional coupling originating from antenna effect. We estimate the coupling efficiency and directivity of the light emitted into the waveguide. \end{abstract}
%
% Uncomment for keywords
%\vspace{2pc}
%\noindent{\it Keywords}: XXXXXX, YYYYYYYY, ZZZZZZZZZ
%
% Uncomment for Submitted to journal title message
%\submitto{\JPA}
%
% Uncomment if a separate title page is required
%\maketitle
% 
% For two-column output uncomment the next line and choose [10pt] rather than [12pt] in the \documentclass declaration
%\ioptwocol
%

\section{Introduction}

In the past decades, strong efforts have been put towards convergence of integrated photonics and quantum technologies \cite{Silverstone_Bonneau_2016,Couteau:Nat23}. 
Notably, linear optics quantum computation (LOQC) requires triggered single-photon sources, which supply photons into photonic circuit \cite{Kok:07}. Recently, several groups have achieved direct on chip integration of single photon source \cite{Kennard:13,Benson:19,Zwiller:20,Kumar_Bozhevolnyi_2021,Humbert_Wiecha_Colas2022}. 
{Realizing non classical light source working at room temperature is a hot topic and colloidal QDs are among the candidates thanks to their engineerable optical properties \cite{Ye:23}.} Moreover, since quantum operation requires photons source presenting a long coherence time, Purcell-enhanced single photon sources are generally used so that the emission dephasing rate becomes negligible \cite{Koenderink:19}. {Purcell effect of a photonic mode is proportional to the mode quality factor and inversely proportional to the mode volume.}  And most of on chip quantum circuits rely on single photon sources coupled to a micro-optical cavities of high quality factor \cite{Senellart:13,Arcari_Song_Stobbe_2014,Holleczek_Brien_Poulios_2016,Lodahl:20}. Another approach is to take benefit from the low mode volume of surface plasmon, pushing further integration capabilities \cite{Wang:16,Weeber-Dubertret:2016,GCF-Barthes-Girard:2016,Shalaev:20}. In this context, optical nano-antennas are promising candidates which provide both Purcell enhanced photon emission and directivity at the same time \cite{Curto_Volpe_2010,grosjean2013optical,Greffet:Patch13,Peschel:13,Ho:18}. Photonic waveguide decorated by optical antenna  strongly focus guided mode onto a subwalength area compatible with single photon source excitation \cite{Bernal_Koenderink_2012,Muskens:18,blanquer2021waveguide}. In this work, we investigate a reciprocal configuration, namely nano-antenna assisted quantum dot (QD) emission into high index planar waveguide. We consider a gold Yagi-Uda plasmonic antenna deposited onto a titanium dioxyde (TiO$_2$) planar waveguide. Yagi-Uda geometry has been chosen for efficient in plane directivity and TiO$_2$ material is a promissing alternative to silicon nitride for applications involving  sources emitting in the visible range \cite{Mazur:12,Weeber-Dubertret:20}. {Since on chip coupling efficiency process is governed by local excitation of the nano-optical antenna, we consider few-photons sources based on CdSe/CdS QDs for characterizing and designing the on-chip plasmonic antenna platform.}

\section{Design of Yagi-Uda antenna}

\begin{figure}[!h]
\centering
\includegraphics[width=0.95\textwidth]{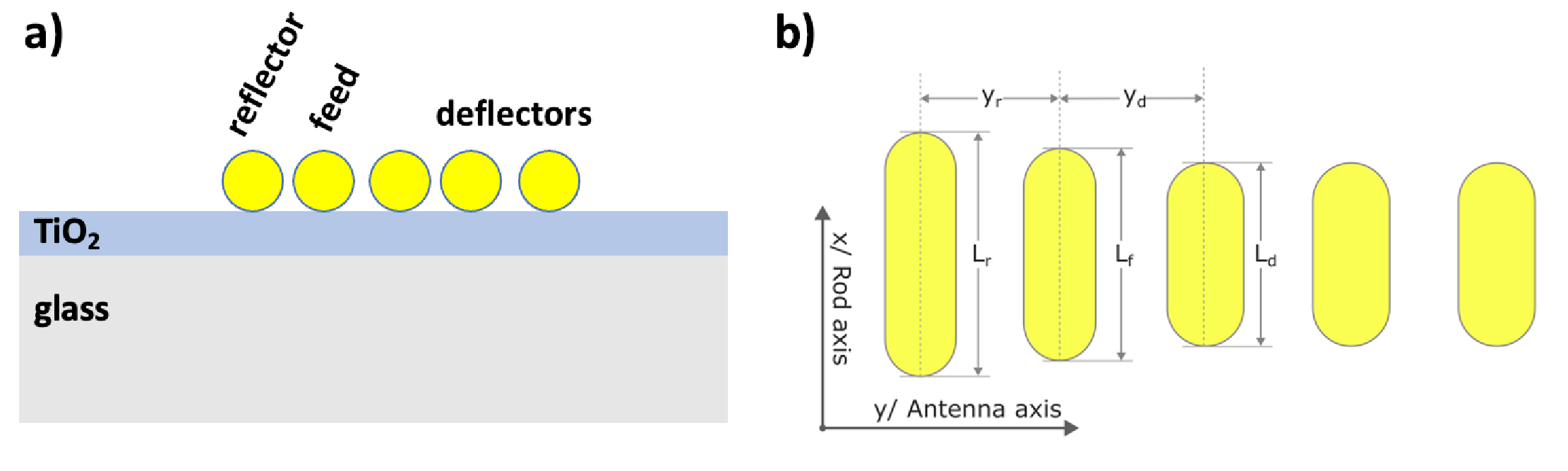}
\caption{Scheme of the Yagi-Uda antenna on a TiO$_2$ planar waveguide. a) Side-view detailing of antenna elements.  b) Top view with dimensions.}\label{Fig:yagi-uda}
\end{figure}

The design rules of Yagi-Uda antenna have been extensively studied in the radio frequency (RF) range \cite{balanis2015antenna} and scale law approaches have been proposed for down-scaling to nano-optics \cite{novotny2007effective,hofmann2007design,bonod2010ultracompact}.	 We  emphasize that plasmonic inverse design via evolutionnary optimization has recently converged towards a Yagi-Uda like configuration, revealing it is close to the optimal \cite{Wiecha:2019}. Yagi-Uda antenna consists of five gold nanorods: a reflector, a feed and three directors, as shown in Fig. \ref{Fig:yagi-uda}. The feed element resonates with the light emitted by a nearby QD. The resonant wavelengths of the reflector and directors are red and blue shifted so that their interaction with  emitted light results in a phase difference and constructive interferences enhance forward scattering whereas destructive interferences decrease backward scattering, leading to high antenna directivity. The lengths of the antenna elements (feed, directors and reflector) govern the antenna resonance whereas their spacing govern the antenna directivity. Rough dimensions of Yagi-Uda antenna are firstly estimated by semi-analytical approach (\S\ref{RFdesign}), which can save simulation time.  The design is then optimized by finite element method (FEM) in section \ref{sec:simu-def}. 
\subsection{Design rules}
\label{RFdesign}
	
\subsubsection{Antenna resonance}
According to antenna theory, the feed length should be half the effective wavelength of the mode propagating along the rod to support a dipolar resonance.  Therefore, we first estimate the effective wavelength $\lambda_{spp}$ of the surface plasmon polariton (SPP) propagating along an infinite nanowire deposited on a TiO$_2$ planar waveguide at the QD emission wavelength $\lambda_0=670$ nm. Parameters are listed in table \ref{tab:paramYU}, leading to an effective index $n_{spp}=3.64$ so that Fabry-Perot like resonances for finite rod length L occurs for $L+2\delta = \lambda_{spp}/2= \lambda_{0}/(2n_{spp})$, where $\delta$ refers to the field penetration depth into air \cite{Agha-GCF:2014}. Penetration depth is negligible for high effective index so that resonance feed rod length is $L_f\approx \lambda_{0}/(2n_{spp})\approx 92$ nm. 
The length of reflector ($L_r$) and director ($L_d$) are deduced  as \cite{taminiau2008enhanced,hofmann2007design}
$L_{r}=1.25L_{f}\approx 115$ nm and $L_{d}=0.9L_{f}\approx 83$ nm, respectively.

\begin{table}
\begin{tabular}{|l|ll|}
\hline
\text{gold nanowire} &  $n_{Au}=0.138+i3.79$ & radius 20 nm\\
TiO$_2$ slab & $n=2.375$ &  thickness 80 nm \\
glass substrate & $n=1.5$ &  \\
\hline
\end{tabular}
\caption{Fixed parameters for the Yagi-Uda antenna design at the QD emission wavelength $\lambda_0=670$ nm. The optical index of gold is taken from bulk tabulated data \cite{Johnson-Christy:1972}.}
\label{tab:paramYU}
\end{table}

\subsubsection{Antenna Directivity}
 For a directivity into the guided mode along the antenna y-axis (diffracted wavevector $\vec{k}=k_{wg}\vec{u_y}$), spacings of feed to reflector ($y_r$)  and feed to directors  ($y_d$)  are estimated from the RF rules as 
\begin{eqnarray}
y_r=\lambda_{wg}/4.4=\lambda_{0}/(4.4n_{wg})=\SI{89}{\nano \meter}, \label{Eq:yr-free} \\
y_d=\lambda_{wg}/4=\lambda_{0}/(4n_{wg})=\SI{97}{nm}, \label{Eq:yd-free}
\end{eqnarray}
where $n_{wg}=1.72$ is the effective index of the TE guided mode, calculated for a 80 nm thick TiO$_2$ planar waveguide lying on a glass substrate. The thickness has been chosen to obtain a single mode waveguide.

\subsection{Optimization of the antenna design}
\label{sec:simu-def}

\begin{figure}[!h]
\centering
\includegraphics[width=8cm]{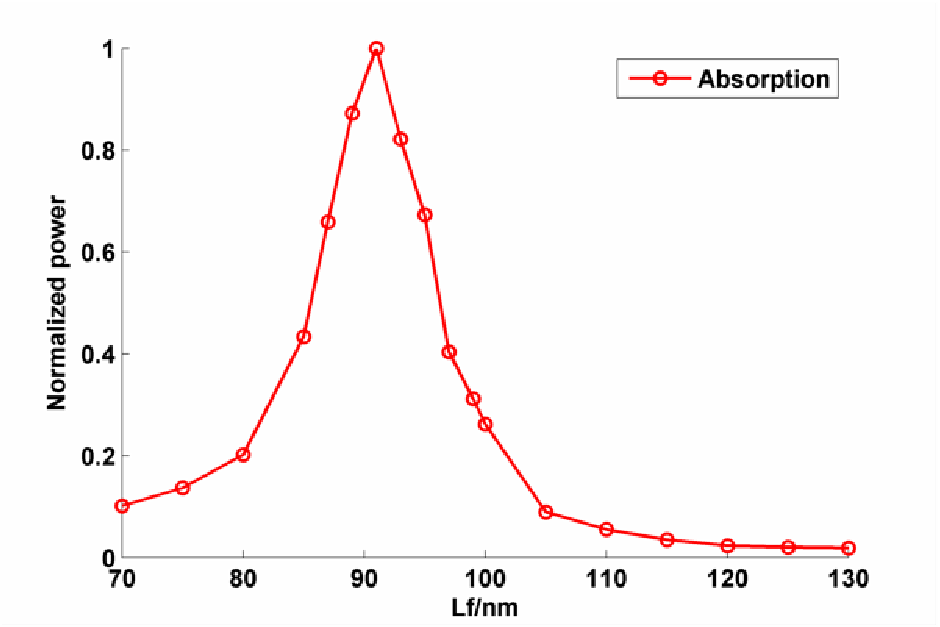}
\caption{Normalized absorption power of single nanorod on a 80 nm TiO$_2$ slab, as a function of the rod length. The rod is excited by a dipole located 5 nm from its extremity.}
\label{Fig:yu-wg-Lf}
\end{figure}
\begin{figure}[!h]
\centering
\includegraphics[width=0.9\textwidth]{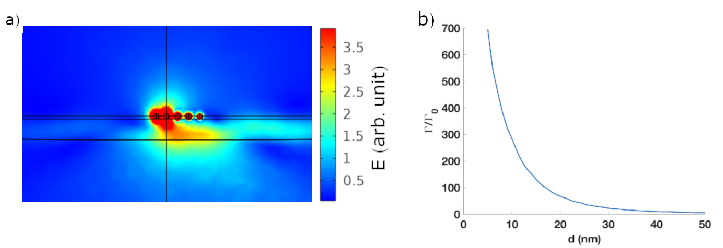}%img2/optimized-Enorm.png}
\caption{a) Electric field norm distribution for a single dipole excitation $p=p\vec{x}$ coupled to Yagi-Uda antenna. The dipole is located 5 nm away from the feed rod. b) Enhancement of the total decay rate as a function of the distance to the feed.}\label{Fig:optimized-corss-section}
\end{figure}

We complete the semi-analytical analysis with full electromagnetic numerical simulations based on finite element method (FEM) \cite{COMSOL}. QD emission is generally modeled by two orthogonal incoherent oscillating dipoles \cite{Buil:2020}, however, the coupling efficiency is dominated by a dipole oriented along the feed rod \cite{taminiau2008enhanced} so that we simplify the study to a single dipole oscillating along the x-axis and at a distance of 5 nm from the feed extremity. Resonant conditions are first determined estimating the time-averaged power absorbed by the Yagi-Uda antenna $<P_{absorption}(\omega)>$, see Fig.\ref{Fig:yu-wg-Lf}. The maximum absorption  is achieved for $L_{f}=92$ nm as expected. We also optimize the length and the position of reflector and deflectors. We systematically sweep these parameters in vicinity of the RF-antenna based model values obtained in \S\ref{RFdesign}. Table \ref{Tab:antenna-num-tio2} gathers the optimized dimensions, in fair agreement with the first estimate of the RF model. Fig.\ref{Fig:optimized-corss-section}a) presents a cross section of QD-TiO$_2$ waveguide coupled emission mediated by the optimized Yagi-Uda antenna. The light scattered into the waveguide, air and substrate can be clearly observed. The light is confined into the TiO$_2$ waveguide and the light power propagates mainly towards the right side, as expected. We also compute the decay rate enhancement in Fig. \ref{Fig:optimized-corss-section}b) and observe two orders of magnitude increase at small distances. This high Purcell factor would lead to improve the indistinguishability of the photons injected into the photonic chip and could be of strong interest for realizing on chip quantum gates at cryogen free temperature \cite{Koenderink:19}. However, this factor includes both radiative and absorption channels so that is has to be manipulated with care. %{Numerical simulations estimate the absorption losses to 50\% for QD in the near-field zone of the feed \cite{taminiau2008enhanced} but this can be significantly reduced considering bowtie or nanoparticle on metal configuration instead of rod for the feed, or dielectric deflectors as in the hybrid metal dielectric antenna of ref. \cite{Ho:18}.} 
Antenna role is also evident from the gain value $G=6.3$ following the RF antenna definition 
\begin{eqnarray}
\label{Eq:gain}
	G=4\pi \frac{P_{max}}{P_{rad}},  \\
	P_{max}=max\{p(\theta,\phi)\}\ \;, P_{rad}=\int_{4\pi str}p(\theta,\phi)  d\Omega 
	\nonumber
\end{eqnarray}
where $p(\theta,\phi)$ is the angular power density of the radiated power and where $P_{rad}$ is the radiation power in full space. For comparison purpose with experimental data, we also estimate the $F/B$ ratio between the right (front - F) and left (back-B) light guided into the slab and obtain {$F/B=4.7=6.7$dB demonstrating a highly directive coupling thanks to the plasmonic nano-antenna: $F/(F+B)=82\%$ of the emitted light is guided to the right. This value is similar to the ratio obtain for free-space antenna or antenna on glass, without any waveguide \cite{Curto_Volpe_2010,Ho:18}, demonstrating the efficiency of the on-chip antenna to redirect QD emission into the waveguide.}

 %as well as antenna characteristics

\begin{table}
\centering
\caption{Optimized dimensions of Yagi-Uda antenna on TiO$_2$ waveguide. }
%\begin{adjustbox}{center}
\begin{tabular}{l|c|c|c|c|c|c|c}\hline
%\backslashbox{Type}{Size} &\makebox[3em]
 & {$L_{feed}$} &\makebox[3em]{$L_{director}$} &\makebox[3em]{$L_{reflector}$}
&\makebox[3em]{$y_{d}$}
&\makebox[3em]{$y_{r}$} 
&\makebox[3em]{$Gain$} 
&\makebox[3em]{$F/B$} \\
\hline
RF model & 92 nm &83 nm & 115 nm & 97 nm & 89 nm & &\\ 
full simulation & 92 nm & 82 nm & 98 nm & 78 nm & 71 nm & 6.3 & 4.7 (in slab)\\
\hline
\end{tabular}
\label{Tab:antenna-num-tio2}
%\end{adjustbox}
\end{table}	

\section{Yagi-Uda on TiO$_2$ microdisk: fabrication and preliminary characterization}	
\subsection{Fabrication of TiO$_2$ planar waveguide}
\begin{figure}[!h]
\centering
\includegraphics[width=8cm]{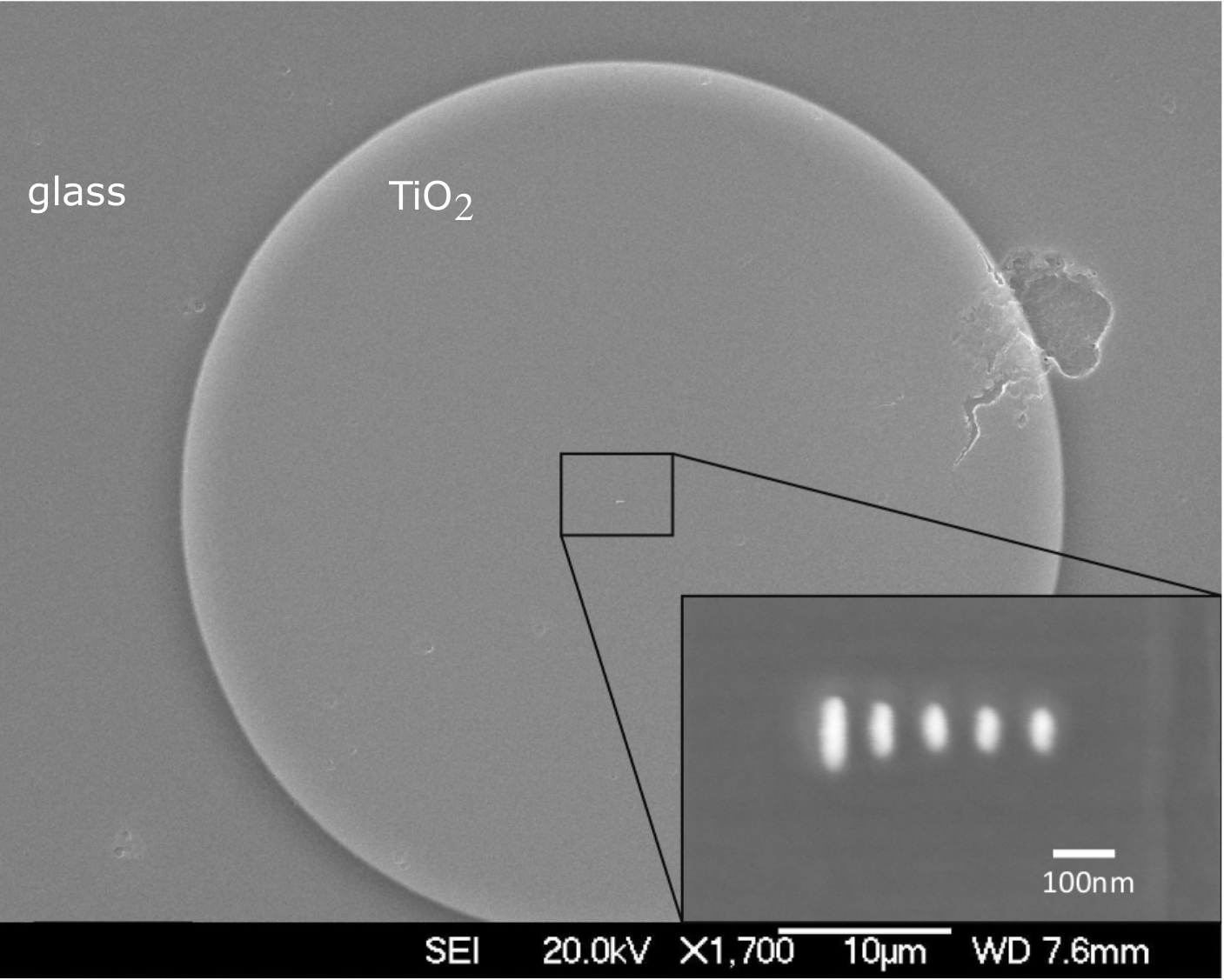}
\caption{SEM image of Yagi-Uda gold nanoantenna centered on TiO$_2$ $ \SI{50}{\micro \meter}$ diameter microdisk planar waveguide. L$_f$=90 nm, L$_d$=85 nm, L$_r$=100 nm, y$_r$=85 nm and y$_d$=90 nm.}\label{Fig:au-on-disk}
%L$_f$=92 nm, L$_d$=83 nm, L$_r$=122 nm, y$_r$=83 nm and y$_d$=90 nm. }\label{Fig:au-on-disk}
\end{figure}

	{On substrate antenna directivity is often determined by Fourier imaging \cite{Curto_Volpe_2010,Huang:2008,Peter:2017}. However, this cannot be applied to bound guided mode of interest for on-chip applications, since no leakage occurs into the substrate.} Instead, we use  $\SI{50}{\micro \meter}$ circular microdisk as planar waveguide so that the optical paths from its center to all directions are the same. The light scattering at the edge helps us to estimate the angular power distribution into the waveguide. We can use it to measure the QD emission directivity. The fabrication involves optical lithography and adheres to the procedures outlined in the literature \cite{hammani2018octave}. The TiO$_2$ layer is firstly deposited on the substrate using physical vapor deposition (PVD). Afterwards, we spin coat positive UV sensitive photoresist AZ MiR 701. The sample is then exposed through a prefabricated photomask which contains microdisk pattern. After baking and development, we use reactive ions etching (RIE) to selectively etch TiO$_2$ not covered by the photoresist. Finally, the sample is immersed in N-Methylpyrrolidone (NMP) solvent to strip the remaining photoresist layer. The Yagi-Uda antenna is fabricated by e-beam lithography. During the process, we use a Cr layer between Au layer and TiO$_2$ to improve the adhesion. After Au deposition and lift-off, we obtain Yagi-Uda antenna on the planar waveguide, as shown in Fig. \ref{Fig:au-on-disk}.
	
\subsection{Preliminary optical characterization}

\begin{figure}[!h]
\centering
\includegraphics[width=14cm]{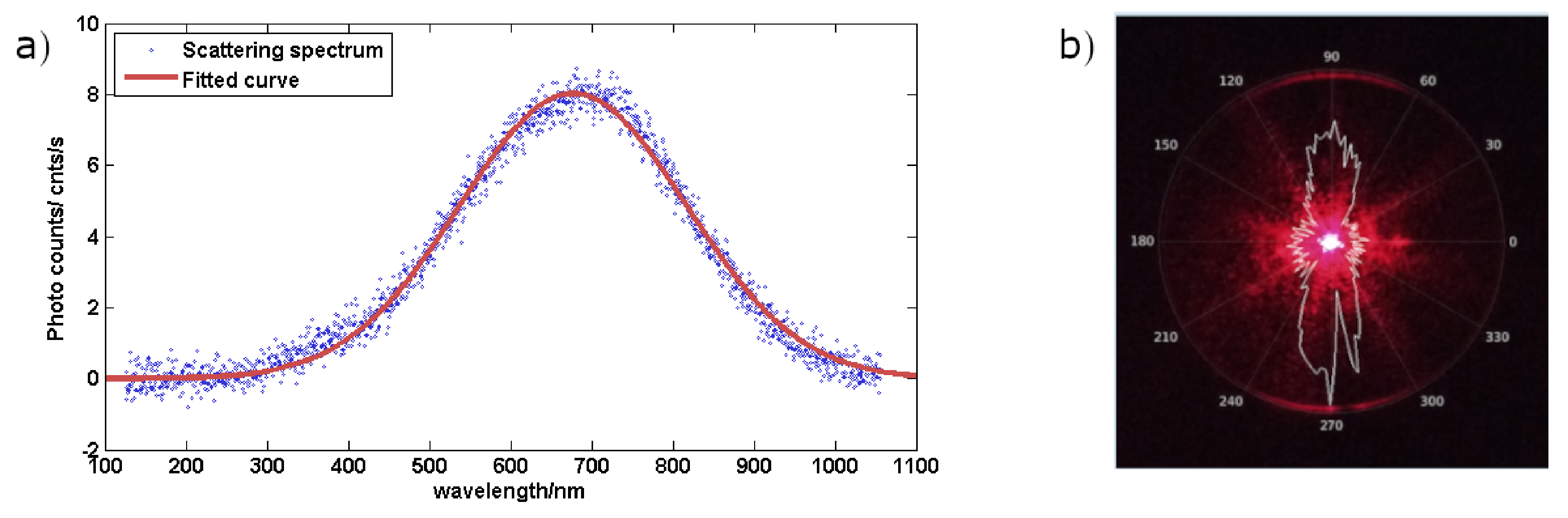}
\caption{a) Dark field spectrum. Spectrum is fitted by Gaussian to estimate the resonance peak. b) Directional light coupling in TiO$_2$ waveguide through Yagi-Uda antenna when excited as a whole with a 670 nm laser light, polarized along the rod (x-axis). Red color is the wide field microscope image of light scattering  and superimposed white line refers to the polar plot of the light scattered along the waveguide ring extremity. 0$^{\circ}$  and 90$^{\circ}$ refer to the antenna y-axis and rod x-axis respectively.}\label{Fig:antenna-scattering}
\end{figure}

We firstly examine the scattering of Yagi-Uda antenna without any local emitter. Dark field spectrum in Fig. \ref{Fig:antenna-scattering}a) shows the antenna resonance close to 670 nm, in line with the designed operating wavelength. In Fig. \ref{Fig:antenna-scattering}, the antenna is excited as a whole by using 670 nm laser to examine its scattering property. The focal size of laser spot is about 300 nm so that all elements are excited simultaneously. The optical microscope image is shown in Fig.\,\,\ref{Fig:antenna-scattering}b). The bright spot in the center is the laser spot illuminating the antenna. The outer ring is the light scattering at the edge of the microdisk. The polar plot superimposed on the image refers to light intensity along the disk perimeter. The light is scattered perpendicularly to the antenna axis, so that no antenna effect is observed, as a consequence of global illumination of the full antenna. In case of local excitation, reflector and directors are excited via the feed element so that phase differences occurs and directivity is expected as observed in the next section. 

\begin{figure}[!h]
	\centering
	\includegraphics[width=0.8\textwidth]{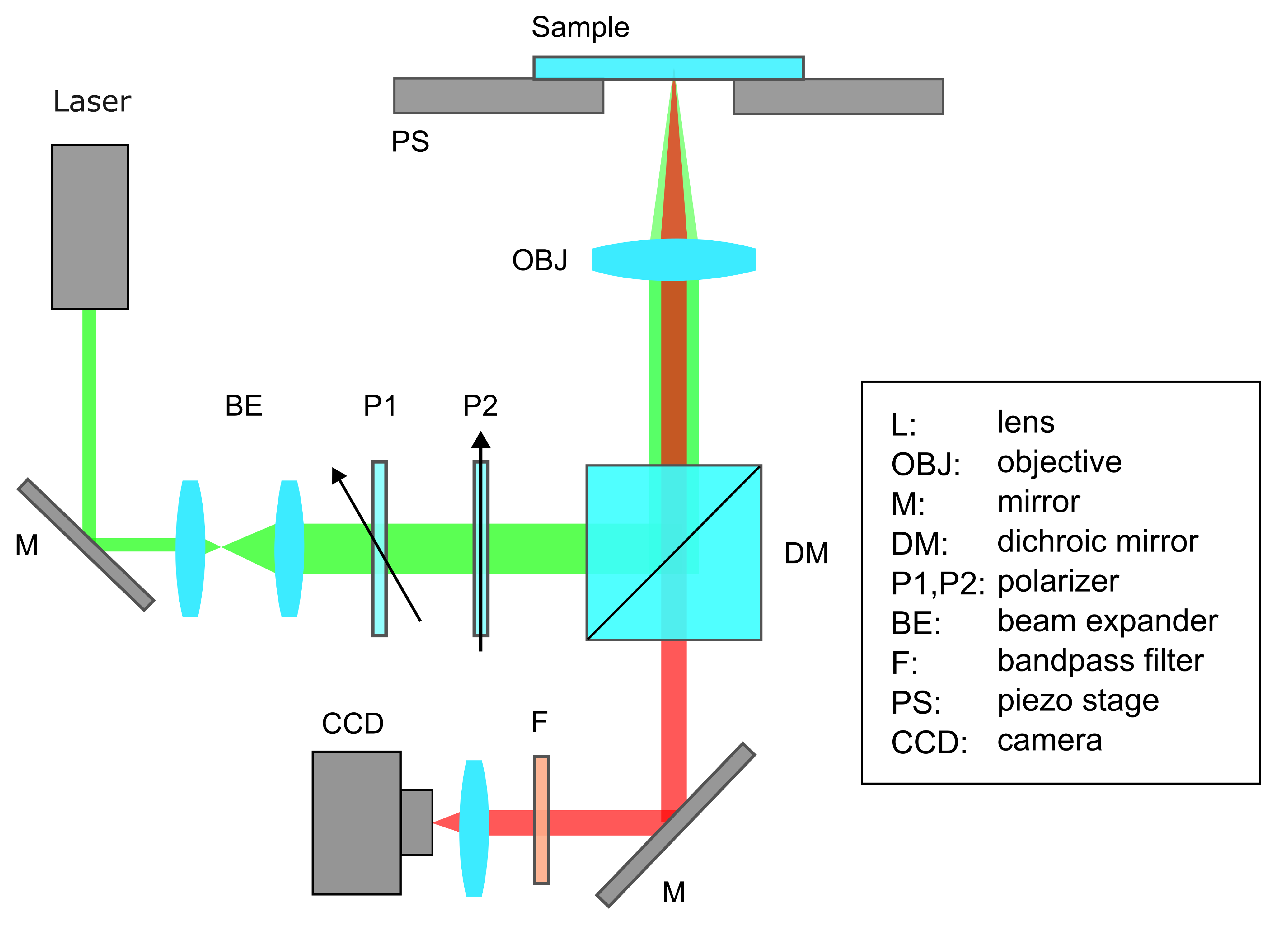}%img2/Fig-ccd-setup.png}
	\caption{Schematic of experimental setup. The QD are excited at the wavelength $\lambda=405$ nm and fluorescence emission at $\lambda=670$ nm is recorded using the same high NA objective (100x, NA=1.49).}\label{Fig:setup}
\end{figure}

\section{Directional QDs emission into planar waveguide}
Finally, we have deposited { CdSe/CdS QDs (Sigma-Aldrich, diameter of 6 nm)} close to the antenna feed, following the method presented in ref. \cite{Weeber-Dubertret:20}. 
QDs are excited at wavelength $\lambda_{exc}=405$ nm and light emitted around $\lambda=670$ nm is detected through the same objective. The scattered light selectively passes through a dichroic mirror and is collected by a CCD camera (see Fig. \ref{Fig:setup}). 

\begin{figure}[!h]
	\centering
	\includegraphics[width=0.6\textwidth]{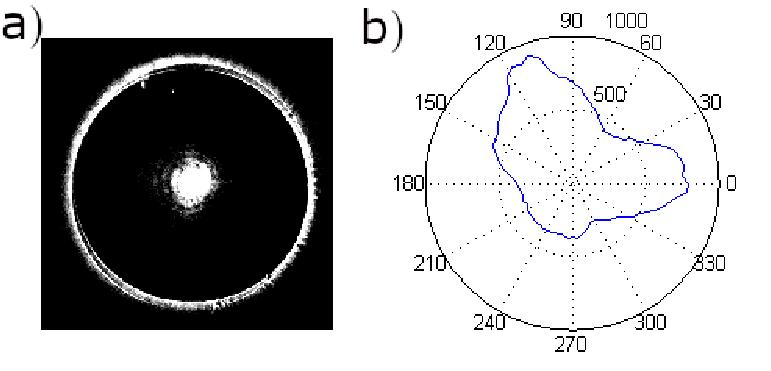}%img2/integrate_final_sum.png}
	\caption{a) Image of scattering light of TiO$_2$ waveguide with Yagi-Uda antenna assisted QD emission. The inner spot is the light scattered by QD and Yagi-Uda antenna. The outer ring is the fluorescence light scattered from the edge of TiO$_2$ waveguide. b) Polar plot of the light scattered along the waveguide periphery.}\label{Fig:integrate-final-sum}
\end{figure}

\begin{figure}[!h]
	\centering
	\includegraphics[width=0.7\textwidth]{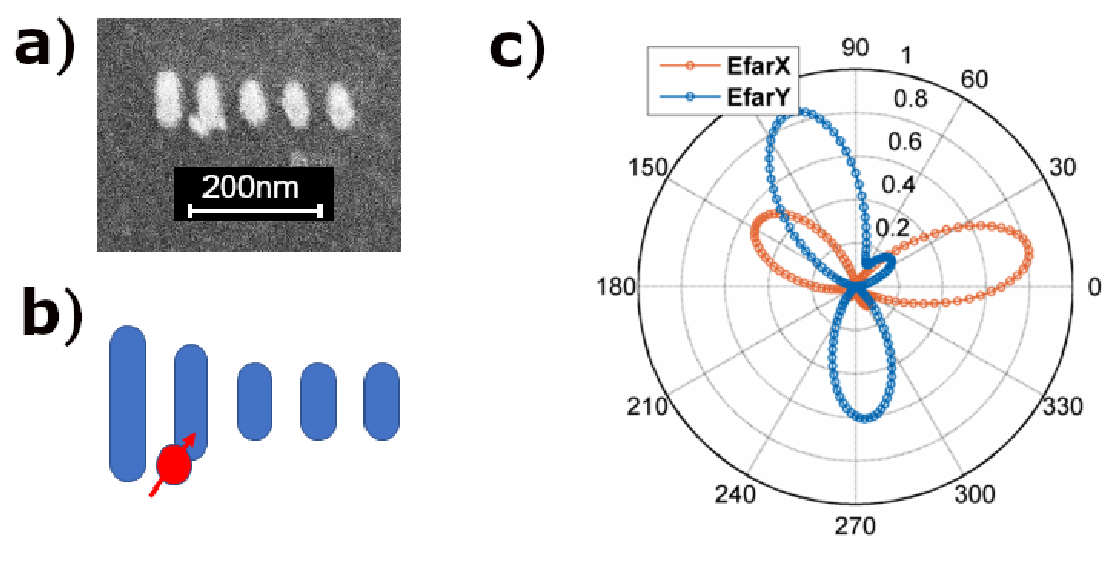}
	\caption{a) SEM image of QDs close to the feed. b) Schemed of the corresponding model. c) Simulated far field polar plots for misaligned coupling.}\label{Fig:final-explain}
\end{figure}

Antenna mediated QD emission coupled to TiO$_2$ waveguide is shown in Fig.\,\,\ref{Fig:integrate-final-sum}a). We estimate the number of QDs below 5 from $g^{(2)}$ autocorrelation measurement (not shown). Fluorescence is collected by a high NA objective. Light coupled to microdisk propagates radially in the waveguide and scatters out at the edge. Since the light source is a local source in the center, its directional property can be estimated by the light scattered out at the periphery of the waveguide. 
\begin{figure}[!h]
	\centering
	\includegraphics[width=0.45\textwidth]{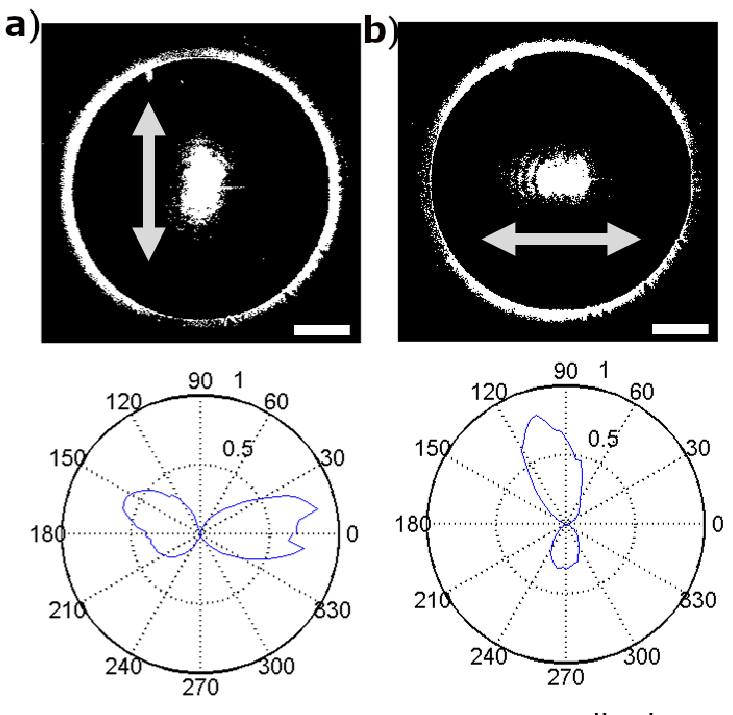}%img2/img-finalV2.png}
	\caption{Antenna mediated QD emission in TiO$_2$ planar waveguide. a,b) Top line: image plane for analyzer parallel (a) or perpendicular (b) to the rod axis. The double arrow refers to the analyser orientation. Bottom line : polar plot of the light scattering at the waveguide ring extremity. Scale bar is $\SI{10}{\micro \meter}$.}
	\label{Fig:img-final}
\end{figure}
We observe directional scattering towards the right (0$^\circ$) in agreement with the antenna effect but also unexpected directionality on the top left side (120$^\circ$). The mode polarization can be determined putting an analyzer before the CCD camera.  In Fig. \ref{Fig:img-final},  we observe that guided light is polarized perpendicularly to the plane of propagation, revealing a TE polarized mode. 
Along the orientation of Yagi-Uda antenna, we measure significant forward directivity with a front-to-back ratio F/B=1.5 (Fig.\,\,\ref{Fig:img-final}a), corresponding to 60\% of the light propagating forward. In the perpendicular direction, we observe additional directivity. We attribute this parasitic directivity to misplaced and or misoriented QDs so that interferences between guided mode directly excited by QD badly coupled to the rod and antenna scattering occurs. SEM image reveals QD slightly shifted to the left of the feed and the far field power computed on Fig.\,\,\ref{Fig:final-explain}a)  presents similar behaviour as experimental datas for both components parallel (x) and perpendicular (y) to the nanorod. The accuracy of QD deposition is estimated to 50 nm, so that it is difficult to deposit QD exactly at the tip of the feed element under the current experiment method. The lift-off process during the QD deposition can also shift the QD position. We expect better coupling with more precise deposition method \cite{Ge_Marguet_2020,dhawan2022fabrication,Humbert_2022}. Probably even higher coupling efficiency could be obtained embedding the Yagi-Uda antenna into the waveguide, thanks to better overlap with the guided mode. However, it worths noticing that {\it a posteriori} selection of a directive signal is still achievable at 0$^\circ$ and 120$^\circ$ directions, of interest for on chip routing.

The coupling efficiency can be estimated from the light scattered around the ring ($P_{ring}$) and in the center ($P_{center}$) ; $\eta=P_{ring}/(P_{ring}+P_{center})=39\%$. It is lower than the calculated one ($\eta_{simu}=64\%$, see \ref{app:simu}), because of lower coupling efficiency depending on QD-feed distance and dipole moment orientation. Numerical simulations considered a single dipole polarized along the rod axis and located at 5 nm from the feed so that it overestimates the coupling efficiency compared to the experimental situation.

\section{Conclusion}
We have investigated Yagi-Uda assisted QD directive emission into TiO$_2$ planar waveguide. Yagi-Uda optical nano-antenna dimensions were first estimated transposing free-space RF antenna concept to on chip waveguide optical nanoantenna that controls emission in a guided mode. Full electromagnetic simulations completed the antenna design. We experimentally measured 39\% coupling efficiency and 60\% forward propagation and numerically demonstrated that up to 64\% coupling efficiency and 82\% forward propagation could be achieved. Taking also benefit of the high Purcell enhancement by the plasmonic antenna, we expect to improve the indistinguishability of the photons injected into the photonic chip, of strong interest for realizing on chip quantum gates at cryogen free temperature. 
The efficiency of the device is very sensitive to the location of the quantum emitters and could be significantly improved thanks to recents development of controlled positioning \cite{Ge_Marguet_2020,dhawan2022fabrication,Humbert_2022}. This hybrid platform based on plasmonic antenna assisted directive emission into photonic waveguide could also be used for lab-on-chip applications combining surface enhanced spectroscopies and photonics capabilities \cite{Baets:16}. It is worth mentioning that location of the emitters is not as critical for ultra compact sensor devices since it can be combined with microfluidic to probe solely the active zone.      

\ack
We acknowledge support from European Union’s Horizon 2020 Research and Innovation Program under Marie Sklodowska-Curie Grant Agreement No. 765075 (LIMQUET), the European Regional Development Fund FEDER-FSE 2021-2027, Bourgogne-Franche-Comté and the French Investissements d’Avenir program EUR-EIPHI (17-EURE-0002). This work has benefited from the facilities of the platforms in Région Bourgogne Franche-Comté SMARTLIGHT (EQUIPEX+ contract ANR-21-ESRE-0040) and ARCEN Carnot. Platform ARCEN Carnot is financed by the Région de Bourgogne Franche-Comté and the Renatech+ CNRS network.

\appendix
%%%%%%%%%%%%%
\section{Simulation of the antenna mediated QD emission into finite planar waveguide}
\label{app:simu}
\begin{figure}[!h]
\centering
\includegraphics[width=\textwidth]{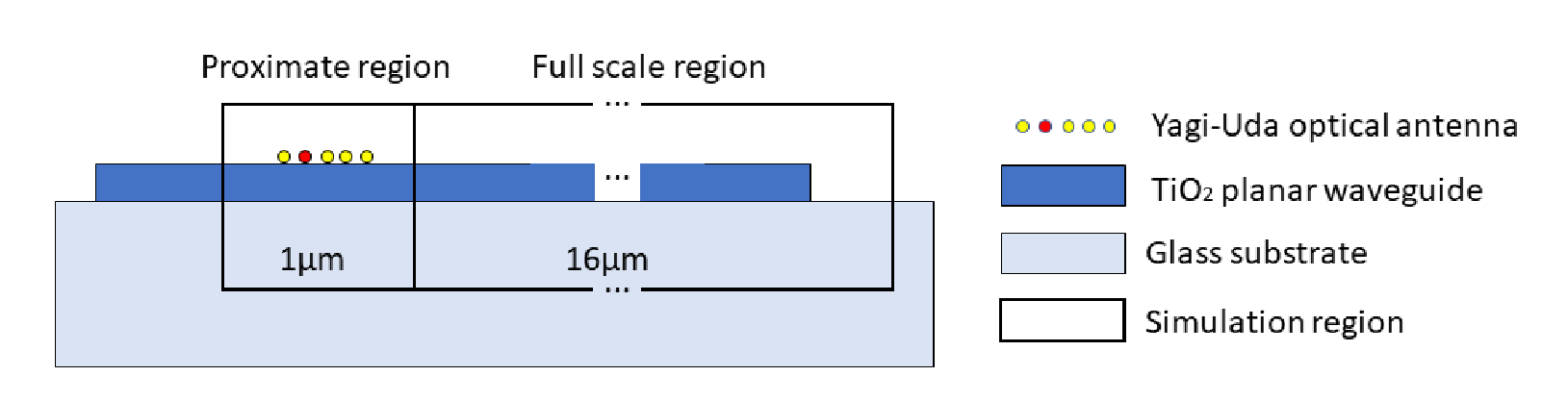}
\caption{Diagram of separated simulation regions.}\label{Fig:simu-region}
\end{figure}

To save calculation time, we separate the simulation work in two steps. A first full 3D simulation is realized as presented in the main text for a Yagi-Uda antenna deposited on an infinite TiO$_2$ planar waveguide (so-called proximate region in Fig.\,\,\ref{Fig:simu-region}).  Then a 2D-simulation permits to characterize scattering at the edge of the waveguide for a modal excitation at the input (so-called full scaled region in Fig.\,\,\ref{Fig:simu-region}). The simulation window of QD-antenna-TiO$_2$ coupling is $\SI{1}{\micro \meter}$ long, and the computational window for investigating guided mode scattering at the edge of TiO$_2$ is $\SI{16}{\micro \meter}$. Finally, we combine the two simulations to estimate the coupling efficiency and compare to the experimental results. 

\subsection{Proximate scale study: Optimized Yagi-Uda antenna on TiO$_2$ infinite waveguide}		
\begin{figure}[!h]
		\centering
		\raisebox{-.5\height}{\includegraphics[width=0.6\textwidth]{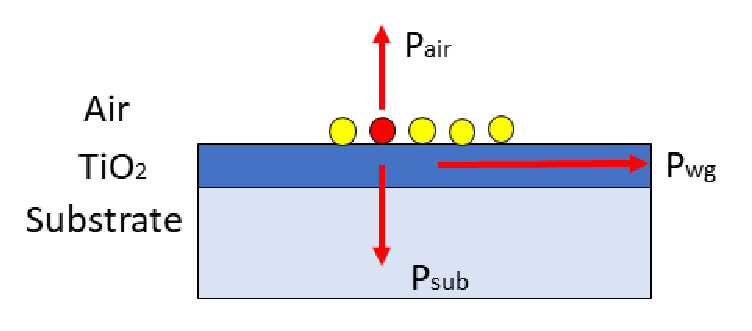}}
		\caption{Three scattering channels of Yagi-Uda antenna on TiO$_2$ planar waveguide in the proximate region.}\label{Fig:proximate-channel}
	\end{figure}

The proximate scale study focus on the optimization of the Yagi-Uda optical antenna. The setup of model is shown in Fig.\,\,\ref{Fig:proximate-channel}. We can distinguish i) QD emission in air (direct or scattering on the antenna, P$_{air}$), ii) fluorescence emission into the substrate (P$_{sub}$) and iii) fluorescence coupled to guided modes (P$_{wg}$). The light power in these three channels will be used to calculate coupling efficiency in the full scale simulation.
	
	\subsection{Full scale study: guided mode propagation and scattering on TiO$_2$ waveguide edges}
In order to compare the simulated coupling efficiency with the experimental measurements, we make full scale ($\SI{16}{\micro \meter}$) 2D simulation to investigate the guided mode propagation and it scattering at the edge of TiO$_2$ waveguide. 	The antenna mediated power P$_{wg}$ coupled to the waveguide is injected into the 2D simulation to estimate the light scattered at the end of the planar waveguide.
	
\subsubsection{2D simulation}
\begin{figure}[!h]
		\centering
		\raisebox{-.5\height}{\includegraphics[width=0.7\textwidth]{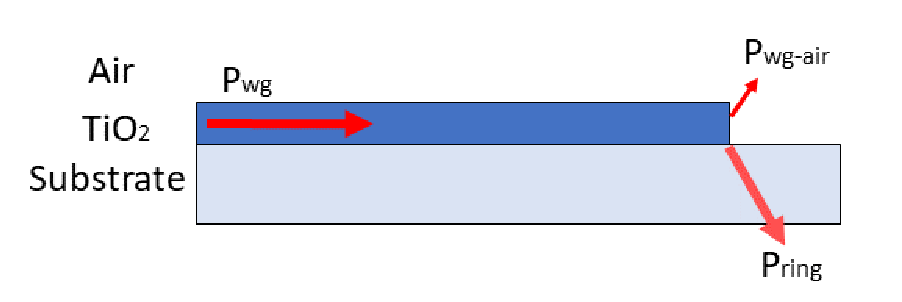}}
		\caption{Three scattering channels of quantum dot fluorescence.}\label{Fig:scatter-channel}
	\end{figure}

	We consider a model with an active input port and guided mode propagation (see Fig.\,\,\ref{Fig:scatter-channel}). The input port emulates the directional light coupled from the proximate scale simulation (P$_{wg}$). The guided mode propagation is regarded as lossless propagation. 2D-simulations are sufficient to characterize the in-out coupling process in the planar waveguide. 
	
\begin{figure}[!h]
		\centering
		\includegraphics[width=0.8\textwidth]{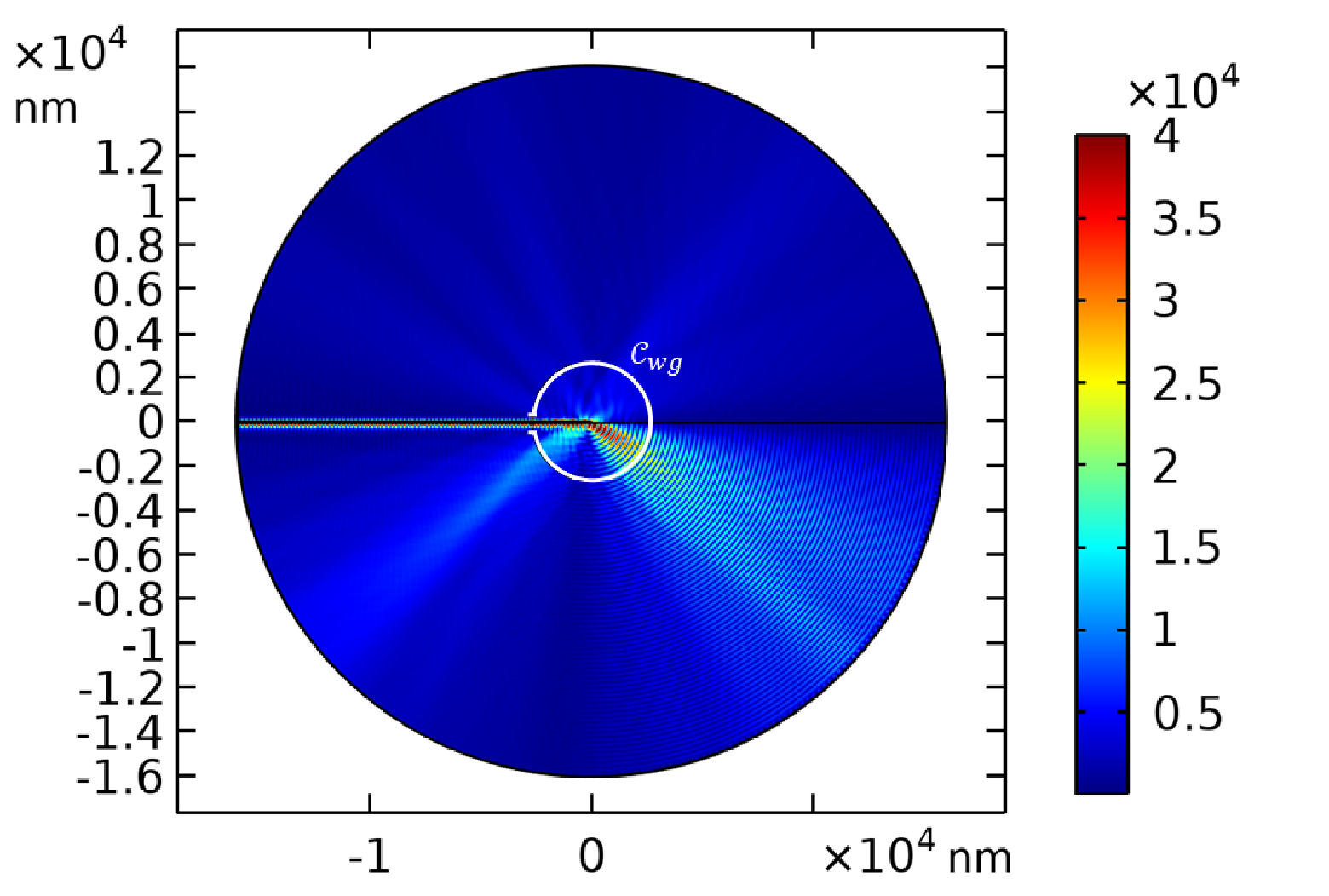}
		\caption{2D simulation of light scattering at the edge of microdisk when excited by a 1W TE mode at the left input. Electric field norm color scale is in V/m. The output power is integrated over the circle $\mathcal{C}_{wg}$ excluding the TiO$_2$ waveguide part.}\label{Fig:tio2-2d-scattering}
	\end{figure}

 A  $\SI{16}{\micro \meter}$ long, 80 nm thick TiO$_2$ waveguide is modeled on glass substrate, see Fig.\,\,\ref{Fig:tio2-2d-scattering}). The waveguide is excited by a numerical port at the left edge of the simulation region. We observe light propagation along the TiO$_2$ waveguide and scattering at the right edge. 
	
\subsubsection{Coupling efficiency}
In the main text, we estimate the coupling efficiency from the light scattered at the output of the planar waveguide and direct QD emission into the substrate 
\begin{equation}\label{Form:experiment-beta}
	\eta=\frac{P_{ring}}{P_{center}+P_{ring}},
\end{equation}
$P_{ring}$ and $P_{center}$ are light power collected by the objective from the end of the waveguide or directly from QD, respectively. 
As we mentioned before, the QD radiative power is divided into three channels: $P_{rad}=P_{air}+P_{sub}+P_{wg}$. 

From the proximate scale model, we estimate the fluorescence light emitted directly from QD ($P_{sub}$) or into the waveguide ($P_{wg}$). The corresponding coupling efficiency are $\eta_{sub} =P_{sub}/P_{rad} = 0.37$ and $\eta_{wg} = P_{wg}/(P_{rad}) = 0.44$. The experimental numerical aperture of the microscope is $N.A.=1.49$, so that the collection angle is $\theta_{col}=arcsin(\frac{N.A.}{n_{imm}})=80^\circ$ where  $n_{imm}=1.51$ refers to the refractive index of the immersion oil. Finally, the collected power is defined by the far-field power integrated over spherical crown with 160$^\circ$ opening angle. 
\begin{equation}\label{Form:coll-eff}
	\eta_{center}=\frac{\int_{-\theta_{col}}^{\theta_{col}} P(\theta)d\theta}{P_{sub}}
\end{equation}
We calculate the collection efficiency of light scattered at the edge of the disk as
\begin{equation}
	\eta_{ring} = \frac{\int_{-\theta_{col}}^{\theta_{col}} P_{scatt}(\theta)d\theta}{\int_{\mathcal{C}_{wg}} P_{scatt}(\theta)d\theta}
\end{equation}
where P$_{scatt}$ is the light scattered at the output and $\mathcal{C}_{wg}$ is a circle excluding the waveguide (see Fig.\,\,\ref{Fig:tio2-2d-scattering}). We obtain $\eta_{center}=0.49$ and $\eta_{ring}=0.74$. 

The final ratio between the scattering collected in the center and at the edge is
\begin{equation*}\label{Eq:final-coupling}
\eta_{simulation}=\frac{\eta_{ring}P_{wg}}{\eta_{center}P_{sub}+\eta_{ring}P_{wg}}=\frac{\eta_{ring}\eta_{wg}}{\eta_{center}\eta_{sub}+\eta_{ring}\eta_{wg}}=64\%.
\end{equation*}
We can reasonably assume that P$_{sub}$ originates mainly from the uncoupled QD emission.  
Therefore $\eta_{simulation}=64\%$ gives a rough estimation of the light coupled to the waveguide.
{This coupling efficiency is the same as for direct coupling to hybrid plasmonic waveguides \cite{Kumar_Bozhevolnyi_2021} revealing that the antenna preserves coupling efficiency while improving directivity and propagation length in low loss dielectric waveguide.}

%%%%%%%%
\bibliographystyle{iopart-num}
%\bibliography{Thesis-mixAdd,/Users/gcolas/Desktop/ADMINISTRATIF/CV/CVtex0/DONNEES/refGCF}
\providecommand{\newblock}{}

\end{document}